\documentclass[a4paper]{jpconf}
\usepackage{graphicx}
\usepackage{subfigure}
\usepackage[mathlines]{lineno}% Enable numbering of text and display math
\usepackage{amsmath} 
\usepackage{wrapfig}

\newcommand{\minerva}{MINERvA}
\newcommand{\pT}{p_\textrm{T}}
\newcommand{\pperp}{p_\perp}
\newcommand{\singlewid}{0.456\columnwidth}
\newcommand{\dalphat}{\delta\alpha_\textrm{T}}

\newcommand{\difd}{\textrm{d}}
\newcommand{\dEdx}{\difd E/\difd x}

\begin{document}

\hfill  FERMILAB-CONF-16-319-E-ND-PPD

\title{Probing nuclear effects using single-transverse kinematic imbalance with \minerva}

\author{X.-G.~Lu$^{1}$ and M.~Betancourt$^{2}$ for the \minerva~Collaboration}

\address{$^{1}$~Department of Physics, Oxford University, Oxford, Oxfordshire, United Kingdom \\
$^{2}$~Fermi National Accelerator Laboratory, Batavia, Illinois 60510, USA}

\ead{Xianguo.Lu@physics.ox.ac.uk}

\begin{abstract}

Kinematic imbalance of the final-state particles in the plane transverse to the neutrino direction provides a sensitive probe of nuclear effects. In this contribution, we report the \minerva~measurement of the single-transverse kinematic imbalance in neutrino charged-current quasielastic-like events on CH targets. To  improve the  momentum measurements of the final-state particles, we develop a  method to select elastically scattering contained (ESC) protons and a general procedure to correct the transverse momentum scales.

\end{abstract}

\section{Introduction}

\begin{wrapfigure}{r}{0.4\textwidth}
\includegraphics[width=0.4\columnwidth]{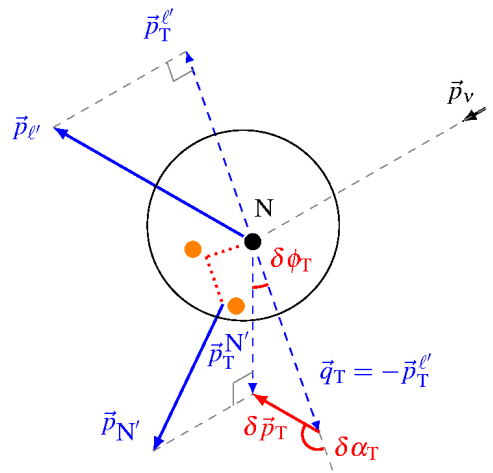}
\caption{Definition of the kinematic variables. For exact definitions, see~\cite{Lu:2015tcr}.}\label{fig:def}
\end{wrapfigure}

Understanding nuclear effects in neutrino-nucleus interactions is crucial for the reconstruction of neutrino energy spectra from which the neutrino oscillation properties are derived. By examining the kinematic correlations between final-state particles in exclusive measurements, a useful equivalence has been established between the nuclear effects and the transverse kinematic imbalance~\cite{Lu:2015tcr}. In this contribution, the \minerva~\cite{Aliaga:2013uqz} measurement with the NuMI~\cite{Anderson:1998zza} low energy (peak at 3 GeV) neutrino beams is presented.  Extending the previous analysis of  CCQE-like events on CH~\cite{Walton:2014esl},  the reconstructed single-transverse kinematic imbalances are studied in simulations. Based on GENIE~\cite{Andreopoulos:2009rq} (Version 2.8.4) results with and without FSIs are compared to demonstrate the physics sensitivity of these new variables.

The event selection requires an interaction vertex in the active scintillator tracker and one muon, at least one proton, and no pion in the final state. The muon is reconstructed by the tracker and the MINOS near detector~\cite{Michael:2008bc}. Protons are identified by the tracker $\dEdx$. Pions are vetoed by cutting on the untracked energy and Michel electrons~\cite{Walton:2014esl}.  Once the final-state momenta are determined, they are projected onto the neutrino transverse plane. In particular, the transverse boosting angle $\dalphat$ is defined as in Fig.~\ref{fig:def}.   In the following sections, improvement in the momentum measurements  beyond the default reconstruction and the achieved sensitivity are  discussed. 

\section{Improvement of proton momentum measurement}
\begin{figure}[!ht]
\begin{center}
%\subfigure[]
{\includegraphics[width=\singlewid]{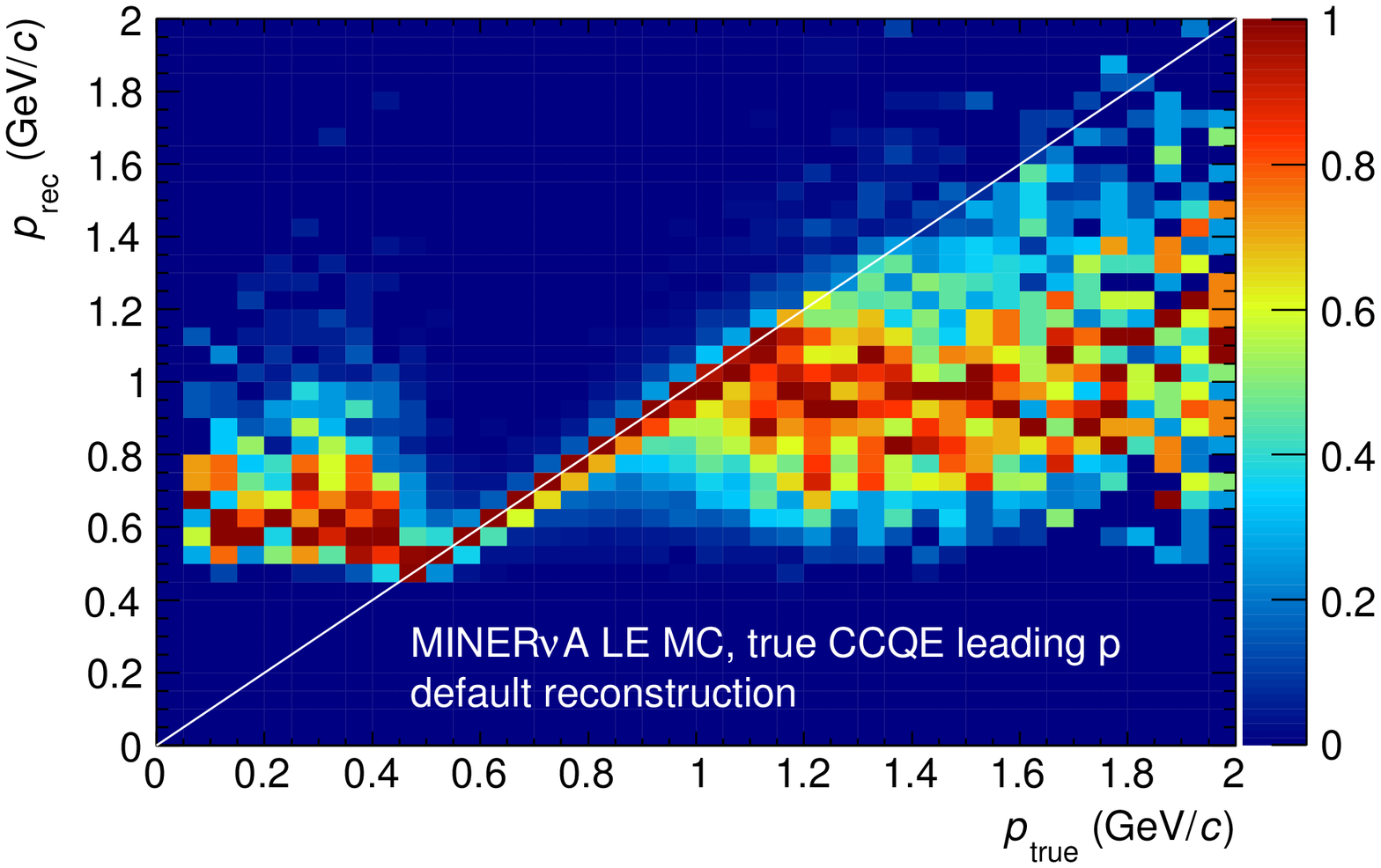}
\includegraphics[width=\singlewid]{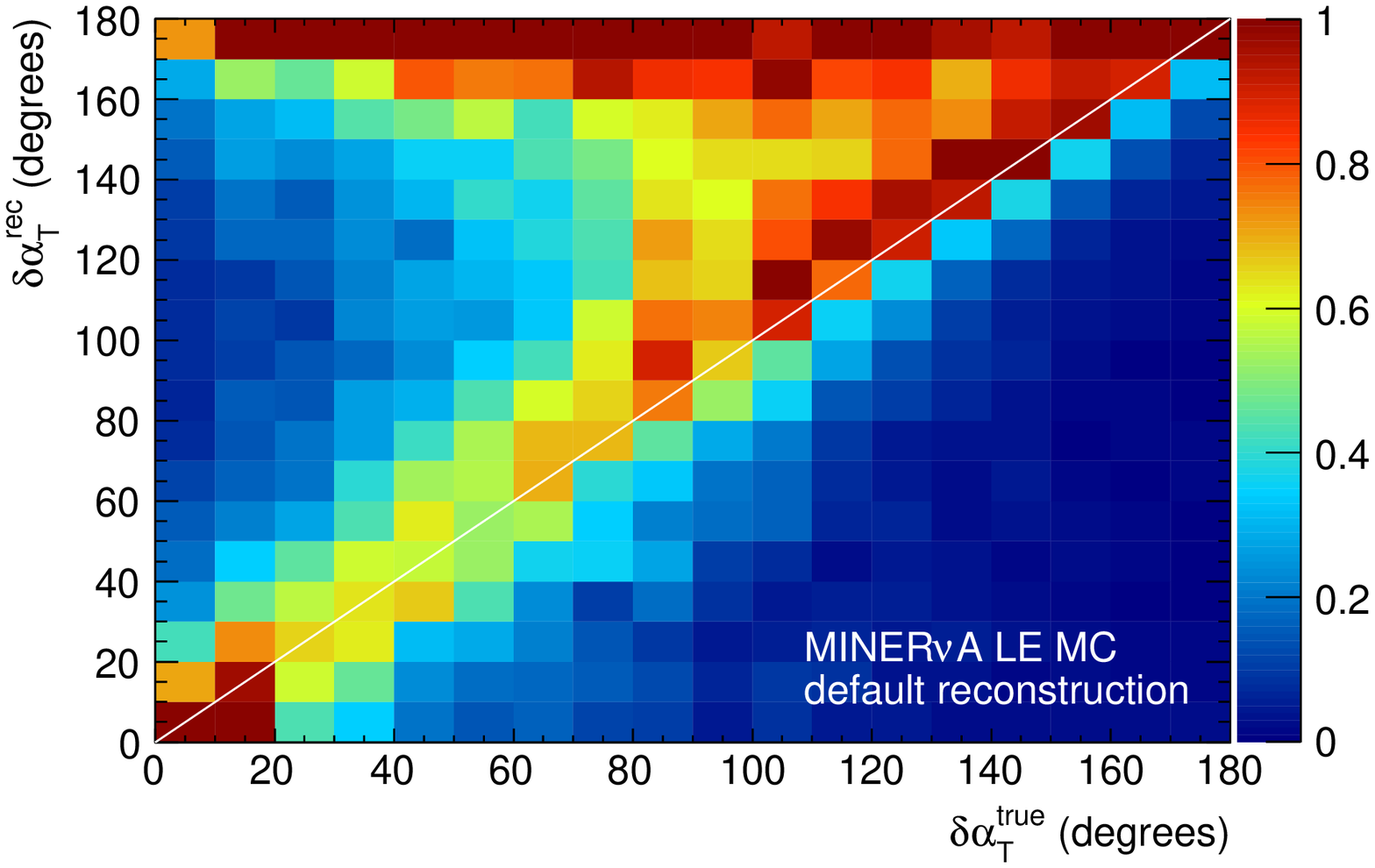}}
%\subfigure[]
{\includegraphics[width=\singlewid]{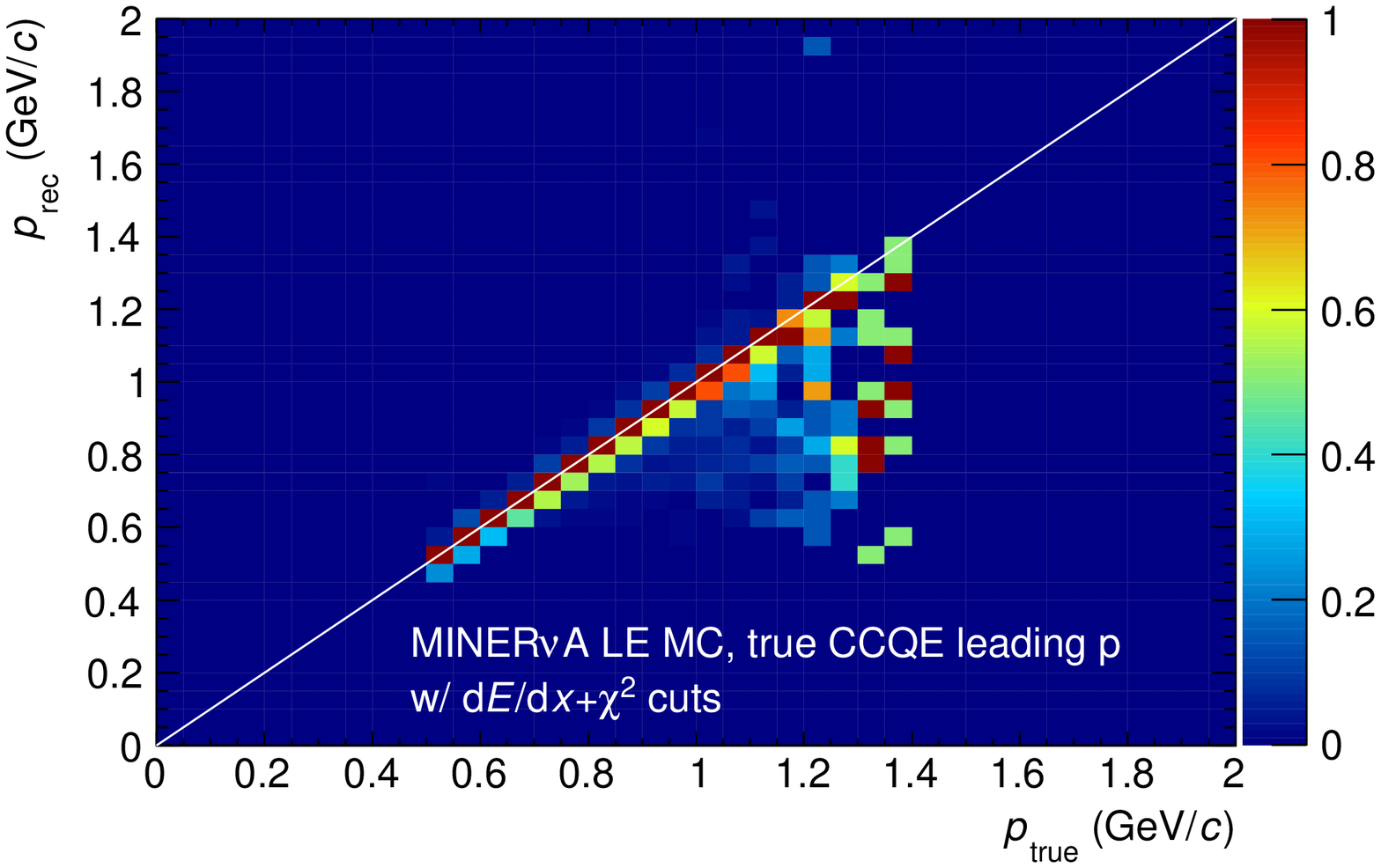}
\includegraphics[width=\singlewid]{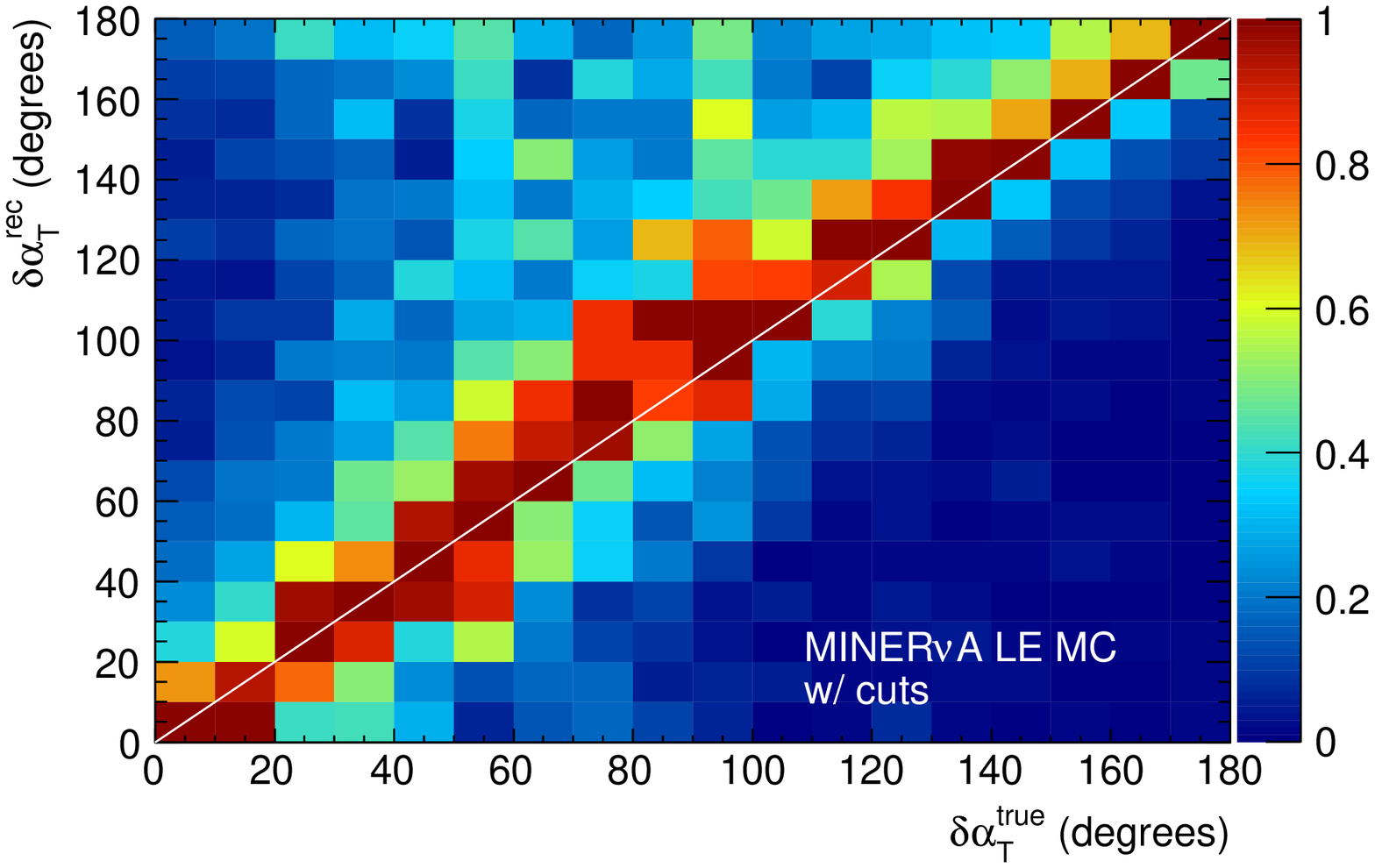}}
\caption{Response matrices by the default reconstruction (\emph{upper}) and  after the $\dEdx$ and $\chi^2$ cuts (\emph{lower}). Each $x$ slice is normalized such that its maximum is 1. }
\label{fig:cuts}
\end{center}
\end{figure}

The proton momentum in the default reconstruction is determined by the $\dEdx$ profile along the track inside the tracker~\cite{Walton:2014esl}. However, for  non-contained or inelastically scattering protons, the profile is sensitive to the non-zero momentum at the track end point and has a distorted  momentum mapping. Elastically scattering contained (ESC) protons, whose end point momentum is 0,  are selected by $\dEdx$ cuts at the last 6 measurement points as they have higher $\dEdx$ near the end point compared to the non-ESC ones. The spread of the reconstructed momentum is reduced to 60\%  of the default value with a  reduction of statistics to 40\%. An additional cut on the track fit  $\chi^2$~\cite{Walton:2014esl}  further reduces the spread to 50\%  and statistics to 30\%.\footnote{With the $\chi^2$ cut alone  the spread and statistics are reduced to 70\% and 60\% of the default values, respectively.}  The improvement on the detector response for the proton momentum and $\dalphat$ is shown in Fig.~\ref{fig:cuts}.  One notes that because non-ESC protons have missing ranges, a momentum measurement by range can also be improved by selecting ESC protons with such $\dEdx$ cuts.

\section{Correction of the proton and muon $\pT$ scales}
\begin{figure}[!t]
\begin{center}
%\subfigure[]
%{\includegraphics[width=\singlewid]{ResolutionCut10_MuonPperpDifflogy0}}
%\subfigure[]
%{\includegraphics[width=\singlewid]{ResolutionCut9_MuonPperpDifflogy0}}
%\subfigure[]
{\includegraphics[width=\singlewid]{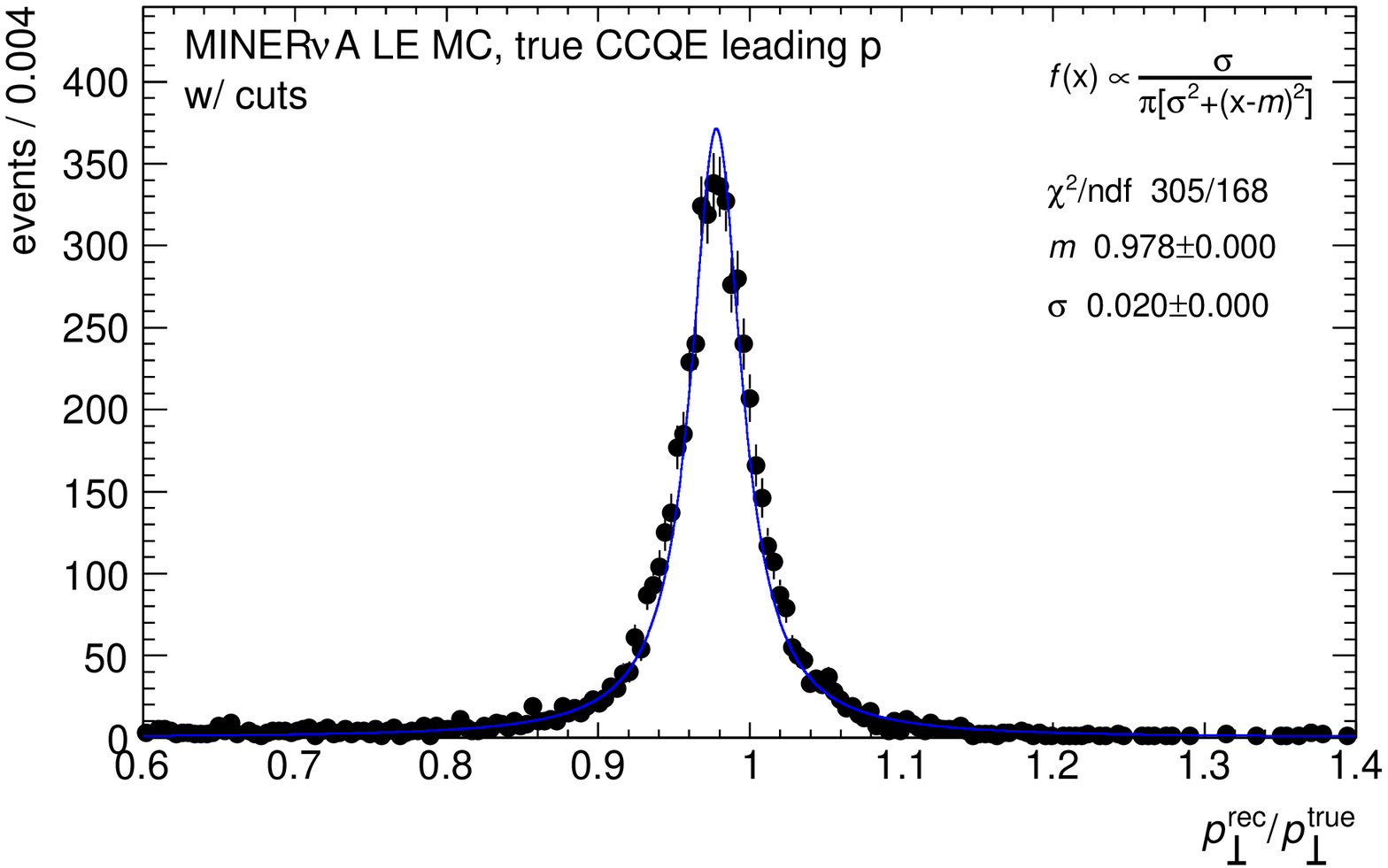}}
%\subfigure[]
{\includegraphics[width=\singlewid]{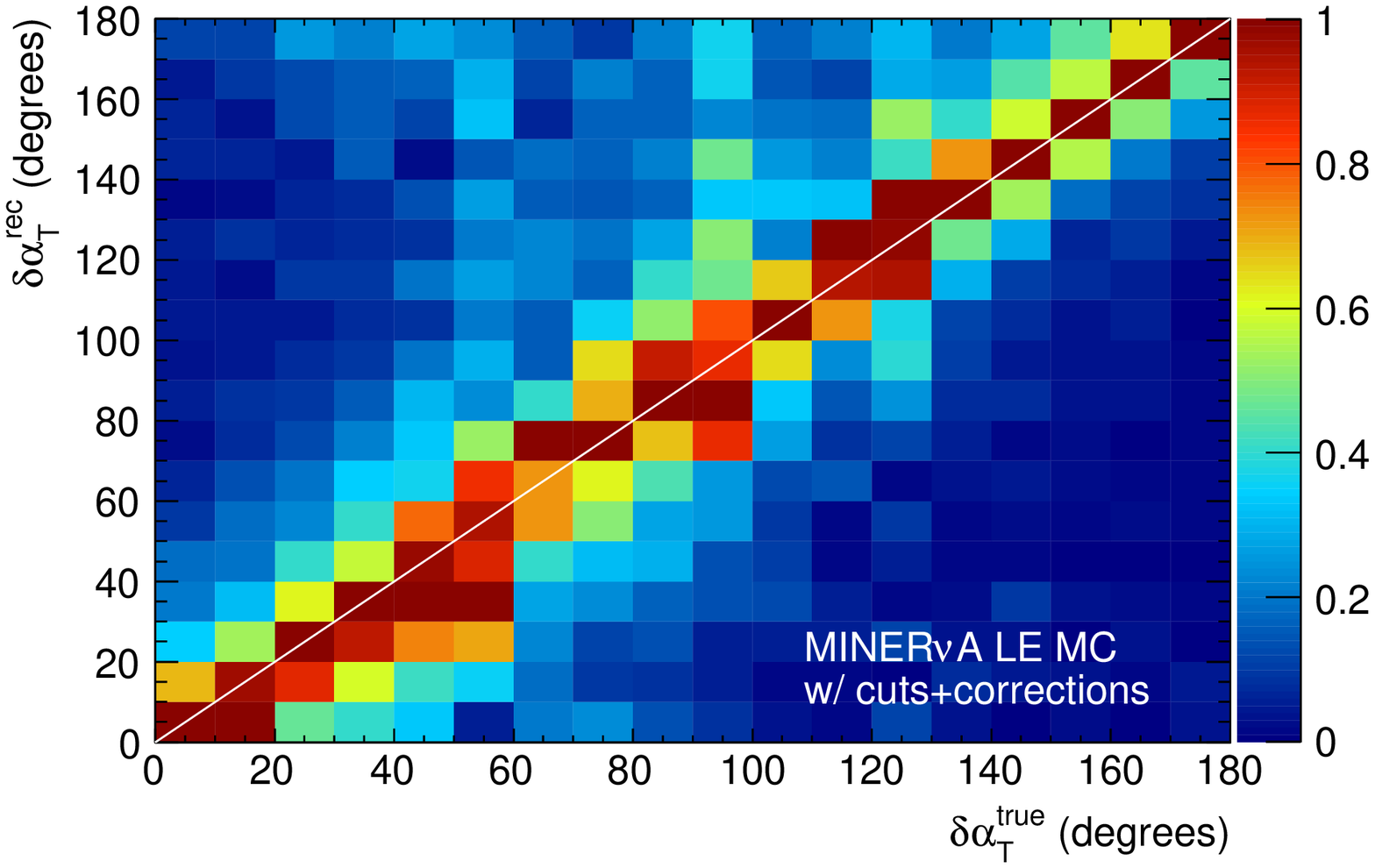}}
\caption{ Distributions of the proton $\pperp$ ratio fitted to a Cauchy function (\emph{left}) and the response matrix for $\dalphat$ after the $\pT$-scale corrections for the proton and muon  (\emph{right}).}\label{fig:corrs}
\end{center}
\end{figure}

Since a non-unity momentum scale leads to an instrumental apparent acceleration or deceleration of the final-state particles, it is crucial to correct the  scales of $\pT$ as the  imbalance is directly  calculated from them. A general recipe  to obtain the scales accurately is to find a $\pT$ scale-related distribution that can be described analytically. In this study,  the distributions of $\pperp^\textrm{rec}/\pperp^\textrm{true}$  for protons and $1/\pperp^\textrm{rec}-1/\pperp^\textrm{true}$ for muons are found to be well described by a Cauchy function ($\pperp$: transverse momentum in the detector coordinate). The fit in the proton case as an example is shown in Fig.~\ref{fig:corrs} (\emph{left}). By these corrections the non-linearity of the detector response is removed [compare Fig.~\ref{fig:corrs} (\emph{right})  to Fig.~\ref{fig:cuts} (\emph{lower right})].

\section{Results}
\begin{figure}[!t]
\begin{center}
%\subfigure[]
{\includegraphics[width=\singlewid]{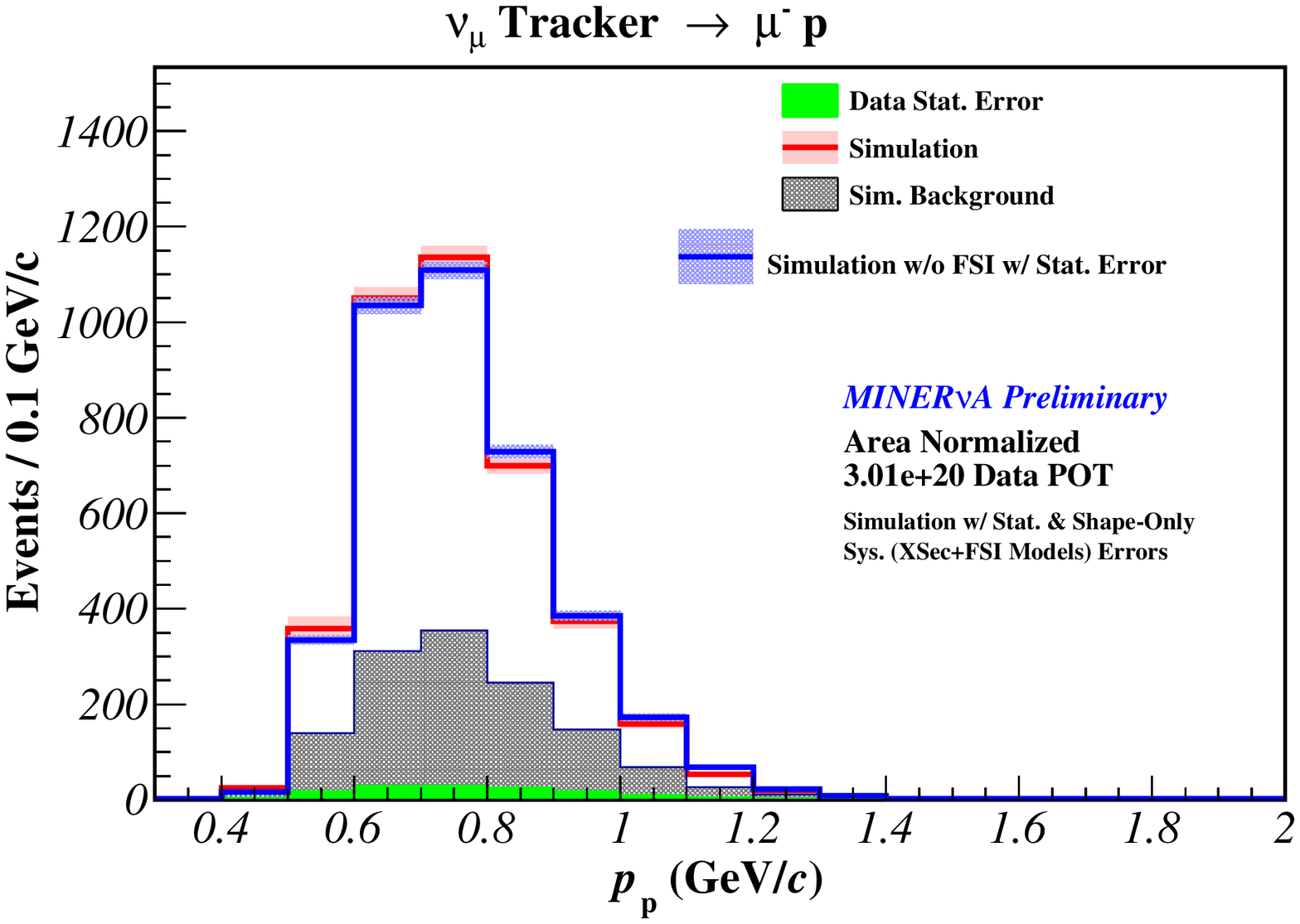}}
%\subfigure[]
{\includegraphics[width=\singlewid]{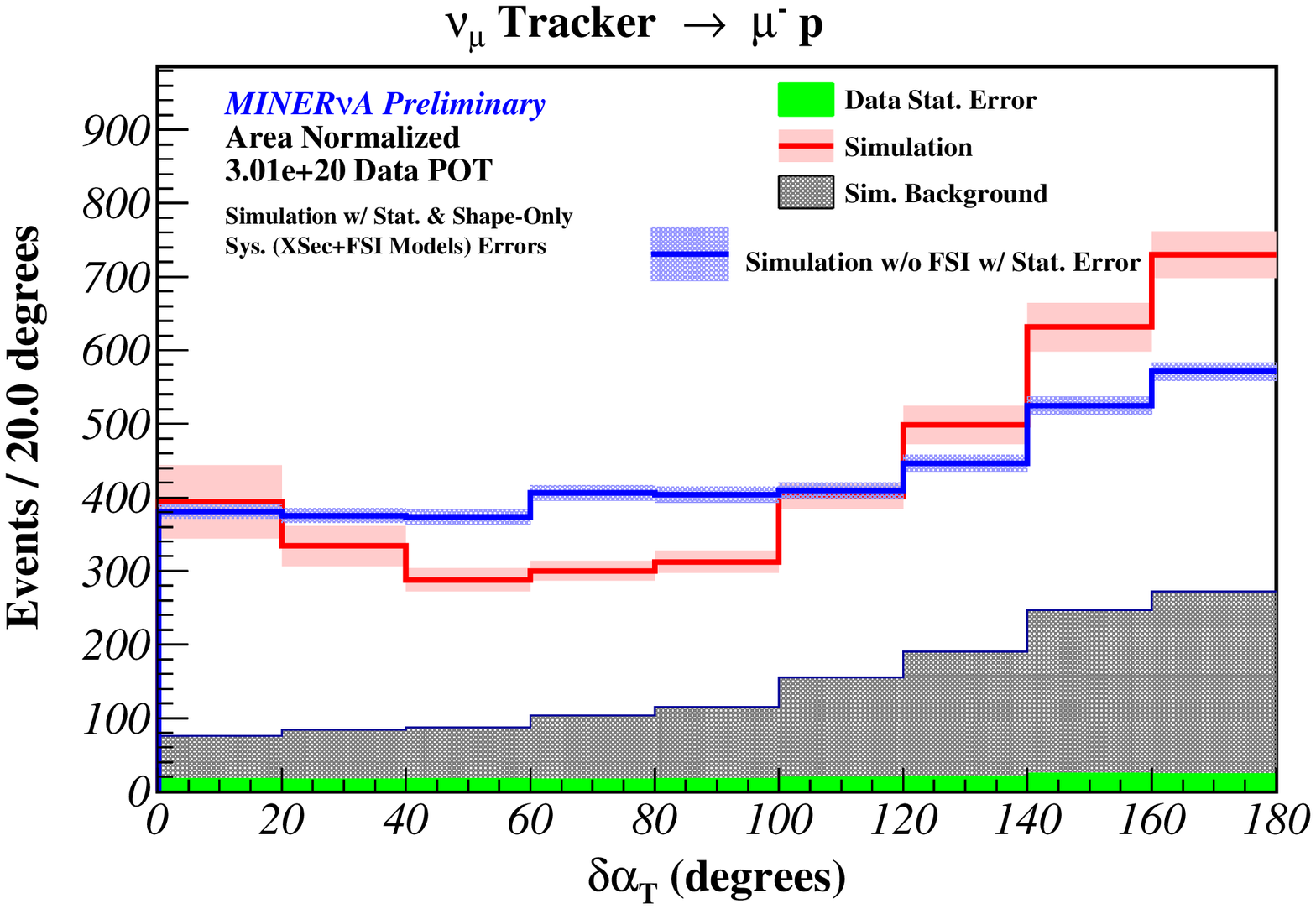}}
\caption{Reconstructed proton momentum (\emph{left})  and $\dalphat$ (\emph{right}) distributions. Simulations with and without FSIs are compared. Resonance production dominates the background. }\label{fig:result}
\end{center}
\end{figure}

As is indicated by Fig.~\ref{fig:result} (\emph{left}), in the given acceptance the overall spectral shapes of the final-state momenta are not sensitive to FSIs. Nuclear effects are  difficult to observe on top of kinematics originating from the neutrino-nucleon interaction level. On the contrary, the distribution of $\dalphat$ clearly shows the effects of FSIs [Fig.~\ref{fig:result} (\emph{right})]. Because of the improvement in the momentum measurement, the peculiar feature of the GENIE ``collinear enhancement''~\cite{Lu:2015tcr}  is  preserved at the reconstructed level. Future comparison to data will examine such predictions.

\end{document}